\documentclass[hyper,
12pt,letterpaper]{JHEP3}

\usepackage{amsmath,amsfonts,amssymb,comment}

\numberwithin{equation}{section}

\def\cN{\mathcal{N}}
\def\cO{\mathcal{O}}
\def\cW{\mathcal{W}}

\def\bR{\mathbb{R}}

\def\SU{\mathrm{SU}}

\def\U{\mathrm{U}}

\def\aa#1#2{a_{#1,#2}}

\newcommand{\ba}{\begin{eqnarray}}
\newcommand{\ea}{\end{eqnarray}}
\newcommand{\nn}{\nonumber}

\newcommand{\Wn}[1]{W^{(#1)}}
\newcommand{\cWn}[1]{\mathcal{W}^{(#1)}}
\newcommand{\wn}[1]{w^{(#1)}}

\def\beq{\begin{eqnarray}}
\def\eeq{\end{eqnarray}}
\newcommand{\nt}{\nonumber\\}

\newcommand{\vev}[1]{ \left\langle {#1} \right\rangle }

\def\cN{\mathcal{N}}
\def\cO{\mathcal{O}}

\def\bR{\mathbb{R}}

\title{$\cN=2$ gauge theories and \\
degenerate fields of Toda theory
}

\let\AA\diamondsuit
\let\BB\heartsuit
\author{Shoichi Kanno$^\AA$,
Yutaka Matsuo$^\AA$, Shotaro Shiba$^\AA$, and Yuji Tachikawa$^\BB$\\

$^\AA$ Department of Physics, Faculty of Science, University of Tokyo,\\
Hongo, Bunkyo-ku, Tokyo 113-0033 Japan\\
$^\BB$ School of Natural Sciences, Institute for Advanced Study, \\
Princeton, New Jersey 08540, USA \\
{\tt kanno,matsuo,shiba@hep-th.phys.s.u-tokyo.ac.jp, yujitach@ias.edu}
}

\preprint{UT-09-25}

\abstract{
We discuss the correspondence between   
degenerate fields of the $\cW_N$ algebra and
 punctures of  Gaiotto's
description of the Seiberg-Witten curve of  $\cN=2$ superconformal
gauge theories.
Namely, we find that the type of degenerate fields of the $\cW_N$ algebra,
with null states at level one,
is classified by Young diagrams  with $N$ boxes, 
and  that the singular behavior
of the Seiberg-Witten curve near the puncture agrees with
that of $\mathcal{W}_N$ generators. We also find 
how to translate mass parameters of the gauge theory to the momenta
of the Toda theory.
}

\begin{document}

\section{Introduction}

Recently, Gaiotto \cite{Gaiotto:2009we} showed that
the S-duality group of 
a large class of four-dimensional $\cN=2$ 
superconformal field theories 
can be understood by realizing them 
by compactifying the six-dimensional $\cN=(2,0)$ theory,
which describes the low-energy dynamics of $N$ coincident M5-branes,
on a Riemann surface $C$ with punctures\footnote{For an extensive review,
see \S 3 of \cite{Gaiotto:2009hg}}.
Each puncture was shown to be labeled by a Young diagram with  $N$ boxes, 
by analyzing the linear quiver gauge theories which fall within this class of theories.
The Seiberg-Witten curve $\Sigma$ of the theory is then given by an $N$-folded
cover of the base $C$: \begin{equation}
x^N + \phi^{(2)}(z) x^{N-2} + \cdots + \phi^{(N)}(z)=0. \label{:SW}
\end{equation} Here, $z$ is a local coordinate of $C$ 
and $\phi^{(k)}(z)$ is a degree-$k$ differential on $C$.
The Young diagram labeling the puncture specifies 
the poles of $\phi^{(k)}$, as will be reviewed later.

Soon, it was demonstrated in \cite{Alday:2009aq} 
by Alday, Gaiotto and one of the authors 
that Nekrasov's partition function of this class of theories with $N=2$
gives the correlation function of the two-dimensional Liouville theory,
which lives on the Riemann surface  $C$ on which the M5-branes are compactified.
There, a puncture on the Riemann surface was identified with
an insertion of the exponential of the Liouville boson.
It was also noted that the ``semiclassical'' limit of  the vacuum expectation
value of the energy-momentum tensor $T(z)$ of the 2D theory gave $\phi^{(2)}(z)$ 
appearing in Eq.~\eqref{:SW}: \begin{equation}
\vev{T(z)}\longrightarrow \phi^{(2)}(z).
\end{equation}

This discussion was then generalized by Wyllard \cite{Wyllard:2009hg} to $N>2$. 
There, the corresponding two-dimensional theory
is the $A_{N-1}$ conformal Toda field theory, 
which is a natural extension of the Liouville theory which is 
equivalent to  the $A_1$ Toda field theory.
The $A_{N-1}$ Toda theory has the $\cW_N$ algebra as its symmetry,
which includes not only  the energy-momentum tensor $T(z)=\Wn2(z)$ 
but also chiral primary fields $\Wn3(z),\ldots,\Wn N(z)$ 
of dimension $3,4,\ldots,N$. 
For $N=3$, Mironov and Morozov \cite{Mironov:2009by} checked 
this proposal up to instanton number 2. 
There have been works to better understand 
why this relation holds \cite{Dijkgraaf:2009pc,Bonelli:2009zp,Alday:2009qq,Nanopoulos:2009uw}.

When setting up the correspondence for general $N$,
it was found by  \cite{Wyllard:2009hg} that
different types of punctures gave different types of states of the Toda theory.
Namely, the ``full puncture'', labeled by the Young diagram  $[1^N]$, 
corresponds to  general Toda momenta, whereas 
the ``simple puncture'', whose diagram is $[N-1,1]$, 
corresponds to a restricted set of Toda momenta and 
gives a degenerate state with  null states in the level-1 descendants. 
However, it has not been understood what are the corresponding states
of the Toda theory for  punctures  labeled by other types of Young diagrams.
Our objective is to answer this question.

The $A_{N-1}$ Toda theory describes, in the minisuperspace approximation, 
a quantum-mechanical wave  in a real space of dimension $N-1$,
scattering off $N-1$ exponential walls.
It is convenient to parameterize the space as\footnote{The authors apologize for the simultaneous usage of $\phi^{(k)}$ as the differentials in 
the Seiberg-Witten curve, and $\varphi_k$ as the bosons of the Toda field theory.
They hope it will not cause deep confusion. $i$ in this paper stands for the imaginary unit.}
$\vec\varphi=(\varphi_1,\ldots,\varphi_N)$ constrained by $\sum \varphi_k=0$.
Then the potential is given by 
$\mu \sum_{k=1}^{N-1} e^{\varphi_k-\varphi_{k+1}}$.
The states correspond to the vertex operator $e^{i\vec \beta\cdot \vec\varphi}$.
We will see that one of the  level-1 descendants becomes null
when $\beta_k=\beta_{k+1}$ for some $k$,
i.e.,~the wave is parallel to the $k$-th wall.
We will then see that the Young diagram $[l_1,\ldots,l_s]$ where $\sum l_k=N$,
specifies the subspace of the momentum  space given by
the form \begin{equation}
\vec \beta=(\underbrace{\beta_{(1)},\cdots,\beta_{(1)}}_{l_1},
\underbrace{\beta_{(2)},\cdots,\beta_{(2)}}_{l_2}, \cdots, \underbrace{\beta_{(s)},\cdots,\beta_{(s)}}_{l_s}).\label{hoge}
\end{equation}

It has been long known that the standard physical states of the Toda theory
have the momenta of the form \begin{equation}
\vec \beta = \vec p-i Q\vec\rho \label{phys:full}
\end{equation} where $Q$ determines the background charge,
$\vec \rho$ is the Weyl vector, and $\vec p$ is a real vector specifying the direction
of the propagation. This never satisfies the condition $\beta_k = \beta_{k+1}$
because of the imaginary part. Instead, we will find a strong indication that 
the physical state of the class specified by a Young diagram $Y$ has the momenta
\begin{equation}
\vec \beta = \vec p_Y - i Q \vec \rho_Y
\end{equation} where $\vec p_Y$ is a general real vector of the form \eqref{hoge}
and $\vec\rho_Y$ a certain real vector determined later in the paper. 
Therefore, the physical states for a generic diagram $Y$ 
are {\em not} a special case of the physical states for the diagram $[1^N]$ given by \eqref{phys:full}.

We will then study the behavior of the differentials 
$\phi^{(k)}(z)$  appearing in  the Seiberg-Witten curve Eq.~\eqref{:SW} 
close to a puncture labeled by $Y$ in the massive gauge theory.
We will see that  their behavior 
agrees with the semiclassical limit of the behavior
of the generators of the $\cW_N$ algebra,
close to the insertion of the  degenerate field of the corresponding type.
In other words, we show that $\phi^{(k)}(z)$ behaves as the semiclassical limit 
of $\vev{\Wn k(z)}$, at least close to the punctures: \begin{equation}
\vev{\Wn k(z)}\longrightarrow \phi^{(k)}(z).
\end{equation}

The rest of the paper is organized as follows.
In \S\ref{rev:SW}, we review the Seiberg-Witten curve of the linear quiver
of $\SU$ gauge groups to remind ourselves how the Young diagram arises
in this setup.
In \S\ref{rev:Walg}, we give a brief review the $A_{N-1}$ Toda field theory
and the $\cW_N$-algebra, including its free-boson realization.
In \S\ref{sec:null}, we construct the null states at level 1,
and show that their structure is characterized by a Young diagram.
We then study how to identify physical momenta for a given diagram,
and also analyze the semiclassical limit of these states.
In \S\ref{sec:SW}, we study the behavior of the Seiberg-Witten 
curve near the punctures. We will see that the null state condition
in the semiclassical limit is exactly reproduced.
We conclude in \S\ref{sec:conclusions}
with a discussion on the future directions.

\section{Review: Young diagrams from Seiberg-Witten curve}
\label{rev:SW}
Let us briefly recall how the punctures 
are labeled by Young diagrams, following the discussion
of \cite{Gaiotto:2009we,Gaiotto:2009hg}; 
readers familiar with this point can skip this section.
Consider a four-dimensional $\cN=2$ linear quiver gauge theory with a chain of $n$ $\SU$ groups
\beq
\SU(d_1)\times\SU(d_2)\times\cdots\times\SU(d_{n-1})\times\SU(d_n), 
\label{eq:linear-quiver}
\eeq
with a bifundamental hypermultiplet between each adjoining gauge group and $k_a$ fundamental hypermultiplets for $\SU(d_a)$. 
To make every gauge coupling constant marginal, the number of the fundamental hypermultiplets must satisfy 
\beq
k_a=2d_a- d_{a+1}-d_{a-1}=(d_{a}-d_{a+1})-(d_{a-1}-d_{a})\,,
\eeq
where we define $d_0=d_{n+1}=0$.
Since $k_a$ is non-negative, we have
\beq
d_1<d_2<\cdots<d_{l-1}<d_l=\cdots=d_r>d_{r+1}>\cdots>d_{n-1}>d_n\,.
\eeq
In the following, we denote $d_l=\cdots=d_r=:N$ and $w_a:=d_{a+1}-d_a$\,.
Since $\sum_{a=0}^{l-1}w_a=d_l=N$, the left tail is characterized by a Young diagram that has a row of width 
$w_a$ for each $a\leq l$. The diagram for the right tail can be assigned in the same way. 
We will denote a Young diagram by listing inside $[\ldots]$ the height of the columns 
in the decreasing order; 
$k$ columns with the same height $h$ are sometimes abbreviated
as $h^k$ inside $[\ldots]$. 
Then, the tail $4>3>2$ corresponds to the diagram $[3,1]$, while
the tail $12>8>4$ is $[3^4]$.
The puncture $[1^N]$ is called the full puncture, corresponding to the tail 
consisting of just one $\SU(N)$ gauge group,
and $[N-1,1]$ is called the simple puncture, corresponding to the tail
$N>N-1>\cdots > 3>2$. Now it is easy to see that there is no distinction
between the simple and the full punctures when $N=2$, because both 
are punctures of type $[1^2]$.

Let us consider the flavor symmetry of the theory. 
A bifundamental hypermultiplet carries $\U(1)$ flavor symmetry. 
It can be associated with a simple puncture.
The flavor symmetry which acts on fundamental hypermultiplets in a tail 
can be read off from the Young diagram, say $[l_1,l_2,\cdots,l_s]$. 
Each column corresponds to a fundamental
hypermultiplet and its height stands for the gauge group it couples to. 
So, if there are $N_h$ columns whose height is $h$,
we have $U(N_h)$ flavor symmetry. Overall $\U(1)$ symmetry is carried by a simple puncture ,
so the flavor symmetry associated with the diagram is
\beq
\mathrm{S}\left( \prod_{h> 0}\U(N_h)\right).\label{eq:flavor}
\eeq

The Seiberg-Witten curve of these linear quiver gauge theories was originally found in \cite{Witten:1997sc}. 
The curve is given by a polynomial of two complex variables $(v,z)$
 of degree $N$ in $v$ and of degree $n+3$ in $z$.
It is sufficient for our further analysis to keep only the left tail in the general form.  
Therefore, for simplicity, we set $l=r=n$ so that we have a full puncture 
on the right hand side of the quiver. The curve is then given by
\beq 
&&\prod_{i=1}^{N}(v-\tilde m_i)z^{n+1}+c_{n}(v^N-M_nv^{N-1}-u_{n}^{(2)}v^{N-2}-u_{n}^{(3)}v^{N-3}-\cdots-u_{n}^{(N-1)}v-u_{n}^{(N)})z^n\nt
&&+\cdots+c_j\left[\prod_{k\,\, \text{s.t.} \, l_k > j}(v-m_k)^{l_k-j}\right](v^{d_j}-M_jv^{d_j-1}-u_j^{(2)}v^{d_j-2}-\cdots-u_j^{(d_j-1)}v-u_j^{(d_j)})z^j \nt
&&+c_0\prod_{k=1}^s(v-m_k)^{l_k}=0.\label{eq:massiveSW}
\eeq
Here, $c_j$ are the gauge coupling parameters and 
$u_j^{(\alpha)}$ parameterize the Coulomb branch. 
$\tilde m_1,\ldots,\tilde m_N$ are the mass parameters 
associated to the  $\SU(N)$ flavor symmetry  of the hypermultiplets
coupled to $\SU(d_n)$, 
$m_k$ controls  the mass of the extra fundamental hypermultiplets 
coupled to $\SU(d_a)$ with $a<n$,
and $M_a$ controls the mass of the  bifundamental hypermultiplets.

It is convenient to rewrite the curve in the following way  \cite{Gaiotto:2009we}.
First, we collect the terms of \eqref{eq:massiveSW} with respect to $v$
and write it as \begin{equation}
\Delta(z)v^N - M(z) v^{N-1} + \sum_{k>1} \psi^{(k)}(z) v^{N-k}=0.\label{eq:auxSW}
\end{equation} 
Then, we perform  the shift \begin{equation}
v\to v-\frac{M(z)}{N\Delta(z)} 
\end{equation} to eliminate the $v^{N-1}$ term,
and finally we change the coordinates from $(v,z)$ 
to $(x=v/z,z)$. The final outcome is 
\beq
x^N+\sum_{k=2}^{N}\phi^{(k)}(z)x^{N-k}=0.
\label{eq:gc}
\eeq
We regard $z$ as a local coordinate of a sphere.
$\phi^{(k)}(z) (dz)^k$ can then be naturally thought of as a degree-$k$ differential.

There are $n+3$ punctures on the sphere. Two of them are at $z=0,\infty$, and 
are labeled by the Young diagrams associated with the tails.
The other $n+1$  are at the zeroes of $\Delta(z)=z^{n+1}+\sum_k c_k z^{k} =0$.
$\phi^{(k)}(z)$  has a pole of order $k$ at each of the puncture;
the Young diagram labeling a puncture is reflected by 
the polynomial relations among the residues of  $\phi^{(k)}(z)$ at the puncture.

The Seiberg-Witten differential is now $\lambda=x dz/z$, and
its residue  at the puncture gives the mass parameters 
of the flavor symmetry associated with it.
It can be read off from Eq.~\eqref{eq:massiveSW}:
\beq
(\underbrace{m'_1,\cdots,m'_1}_{l_1}, \cdots, \underbrace{m'_s,\cdots,m'_s}_{l_s}) 
\label{eq:residue}
\eeq
where $m'_k=m_k-m$ with $m$ appropriately chosen to have $ \sum_{k}l_km'_k=0$.
This pattern agrees with the flavor symmetry \eqref{eq:flavor}.

In the M-theory point of view, this setup describes 
the low-energy effective theory of $N$ coincident M5-branes 
wrapping  the sphere. At the punctures, 
there are  codimension-two defects 
whose types are specified by the Young diagram.

Under the correspondence of the 4D gauge theory and the  2D Toda field theory,   
punctures on the Riemann surface translate into insertions of primary operators. 
As we reviewed in this section, punctures are classified by Young diagrams. 
So it is natural to conjecture that the corresponding primaries 
are also classified by the same diagram. 
We will see below that the structure of the  level-1 null states in $\cW_N$ algebra,
which governs the symmetry of the Toda theory,
is also labeled by the Young diagram with
$N$ boxes,
and that null state conditions  are directly reflected by the behavior of the Seiberg-Witten curve near  the puncture.

\section{Review: $A_{N-1}$ Toda theory and  $\cW_N$ algebra}
\label{rev:Walg}
In this section, we summarize rudimentary materials concerning
the $A_{N-1}$ Toda theory and 
the representation theory of $\cW_N$ algebra.
The standard references are \cite{Fateev:1987zh,Bouwknegt:1992wg};
for the modern developments, see \cite{Fateev:2007ab} and the references therein.
Readers who are familiar with these structures
should proceed to the next section.

\subsection{$A_{N-1}$ Toda theory}
The $A_{N-1}$ Toda field theory is given by the action \begin{equation}
S=\int d^2\sigma \sqrt{g}\left[
\frac1{8\pi} g^{xy} \partial_x\vec\varphi\cdot \partial_y\vec\varphi
+\mu \sum^{N-1}_{k=1} e^{b \vec e_k \cdot \vec \varphi}
+ \frac{Q}{4\pi} R \vec \rho\cdot \vec \varphi 
\right].\label{eq:toda-lagrangian}
\end{equation}
Here $\vec\varphi=(\varphi_1,\varphi_2,\ldots,\varphi_N)$ 
with the condition $\sum \varphi_k=0$ parameterize
the Cartan subspace of the algebra $A_{N-1}$,
$b$ is a real parameter, $Q=b+1/b$,
$e_k$ is the $k$-th simple root of $A_{N-1}$ given by $(0,\ldots,1,-1,\ldots,0)$,
and $\vec\rho$ is the Weyl vector of $A_{N-1}$ given by the condition 
$\vec e_k\cdot \vec\rho=1$ for all $k$.
Explicitly it is given as, $\vec\rho=(\frac{N-1}{2},\frac{N-3}{2},\cdots, -\frac{N-1}{2})$.
This theory is conformal with the central charge \begin{equation}
c=(N-1) + 12 Q^2 \vec\rho\cdot\vec\rho = (N-1) (1 + N(N+1) Q^2).\label{eq:todacentral}
\end{equation}

In the minisuperspace approximation, it describes the propagation of
a quantum-mechanical wave in the space of $\vec\varphi$,
scattered by $N-1$ exponential potential walls perpendicular to $\vec e_k$. 
Therefore, an eigenstate of the Hamiltonian 
roughly corresponds to a linear superposition of 
the waves with different momenta, related by the Weyl reflections 
by the walls. 
In the conformal field theory language, the primary field $e^{i\vec\beta\cdot \vec\varphi}$ 
is known to correspond to the propagating mode with the momenta 
$\vec\beta+iQ\vec\rho$, with $L_0=\vec \beta\cdot \vec\beta/2 + iQ\vec\rho\cdot \vec\beta$.
Two such operators 
$e^{i\vec\beta\cdot\vec\varphi}$ and 
$e^{i\vec\beta'\cdot\vec\varphi}$ are in fact proportional to each other
if \begin{equation}
\vec\beta + iQ\vec \rho = w(\vec\beta' + iQ\vec \rho ) \label{eq:weyl}
\end{equation} for an element $w$ of the Weyl group, i.e.~the reordering of the 
components.  The proportionality constant is called the reflection amplitude 
and was determined in \cite{Ahn:1999dz}. 
The symmetry of the theory is described by the $\cW_N$ algebra,
which will be discussed below.

For $N=2$ with $\vec\varphi=(\varphi,-\varphi)$,
this theory reduces to the standard Liouville theory. 
Let us write $\vec\beta=(\beta,-\beta)$. Then 
$e^{i\vec\beta\cdot\vec\varphi }= e^{2i\beta\varphi}$, 
and the Weyl reflection \eqref{eq:weyl} becomes \begin{equation}
\beta \to -iQ-\beta.\label{eq:liouville-reflection}
\end{equation}

\subsection{$\cW_N$ algebra}

$\cW_N$ algebra is generated by
the energy-momentum tensor $T(z)=\cWn2(z)$
and  $N-2$ chiral primary fields 
$\cWn r(z)$ with $r=3,\cdots,N$, where $r$ also gives the dimension
of $\cWn r$.
The explicit form of the algebra for $N=3$ 
was first determined in \cite{Fateev:1987vh}, 
\ba
\cWn2 (z) \cWn2(0)&\sim & \frac{c}{2z^4} +\frac{2}{z^2} \cWn2(0) 
+ \frac{1}{z} \partial{\cWn2} (0)+\cdots\,,\\
\cWn2 (z) \cWn3(0)&\sim & \frac{3}{z^2} \cWn3(0) + \frac{1}{z}\partial{\cWn3}(0)+\cdots\,,\\
\cWn3(z) \cWn3(0) &\sim & \frac{c}{3z^6} +\frac{2}{z^4} \cWn2(0) + 
\frac{1}{z^3}\partial {\cWn2}(0) + \frac{1}{z^2} \left(\frac{3}{10}\partial^2{\cWn2} (0)+2q^2 \Lambda(0)\right)\nt
&& ~~~+\frac{1}{z} \left( \frac{1}{15} \partial^3{\cWn2}(0) + q^2 \partial\Lambda(0)\right)+\cdots\,,\label{W3}
\ea
where,
\ba
\Lambda = :(\cWn2)^2:-\frac3{10} \partial^2 \cWn2,\quad
q^2=\frac{16}{22+5c}\,.
\ea
The operator $\cWn2$ generates  Virasoro algebra and the operators
$\cWn3,\cdots,\cWn n$ represent the extra symmetries.
The appearance of the nonlinear term $\Lambda$ is a characteristic feature of a
$\mathcal{W}$-algebra. As the dimension of the generators is getting higher,
we have more complicated nonlinear terms in the algebra.


In order to manage the algebra, a representation in terms
of free bosons was developed \cite{Fateev:1987vh,Fateev:1987zh}.
Roughly speaking, this corresponds to taking $\mu=0$ in the Lagrangian \eqref{eq:toda-lagrangian}.
The operators $\Wn r$ are systematically produced through
an $N$-th order differential operator:
\begin{equation}
R_N= \mathopen{:}\prod_{m=1}^N \left( Q \frac{d}{dz} +
\vec h_m \partial_z \vec \varphi\right)\mathclose{:} 
= \sum_k  W^{(k)}(z) \left(Q\frac{d}{dz}\right)^{N-k}\,.
\label{miura1}
\end{equation}
The relation between $\vec\varphi$ and $W^{(k)}$ is called as quantum
Miura transformation.  
In the following, we refer to $R_N$ as ``Lax operator'' for simplicity
by using the terminology of solvable system.
Here, $\vec \varphi(z)=(\varphi_1(z),\cdots, \varphi_N(z))$ are free bosons which
satisfies the operator-product expansion\,: $\varphi_j (z) \varphi_k(0) \sim -\delta^{jk} \log(z)$. 
We write their components as
\ba
\varphi_j(z)=x_j + \aa j0 \log z -\sum_{s\neq 0} \frac{\aa js}{sz^s}\,.
\ea
$\vec h_m$ are vectors in $\bR^N$ and defined by
$(h_j)_{k}=\delta_{jk}-\frac{1}{N}$.
Since it satisfies $\sum_{m=1}^N\vec h_m=0$, 
one component $\sum_k \varphi_k$  of $\vec\varphi$ is decoupled.
 The definition (\ref{miura1}) gives
 $W^{(0)}(z)=1$ and  $W^{(1)}(z)=0$. The Virasoro generator is
\ba
W^{(2)}(z) &=&-\frac{1}{2}:(\partial_z\vec\varphi)^2: + Q\vec\rho\cdot \partial_z^2\vec\varphi 
\ea with the central charge \eqref{eq:todacentral}.
It is important to note that $\Wn k$ defined by the quantum Miura 
transformation is not primary in general. 
For example, a primary $\Wn 3$ is given by \cite{Bouwknegt:1992wg}
\ba
\hat W^{(3)}(z) = \Wn3(z)-  \left(
\frac{N-2}{2}
\right)Q\partial \Wn2(z)\,. \label{eq:primaryW3}
\ea  $\cWn3(z)$ quoted in \eqref{W3} is then 
$\cWn3(z)=i\sqrt{3}q \hat W^{(3)}(z)
$.

The Lax operator (\ref{miura1}) plays an essential role to derive the
relation between the $\mathcal{W}_N$ algebra with the $A_{N-1}$ Toda equation
\cite{Bilal:1988ze}.  It was shown that the Toda fields $\varphi_p$
may be rewritten as the Wronskians of the solutions to $R_N \psi=0$.
Geometrical aspects of the correspondence was given in \cite{Gervais:1992gg}.
Such correspondence will, however, not be explored further in this paper
since we have to combine the left and right mover to give the Toda fields.  
Instead, we will treat
Toda fields $\vec\varphi$ as free fields and treat the Toda potential
as chiral perturbation.

In this free-boson representation, the primary fields are defined by the
vertex operators 
\ba
&& V_{\vec\beta}(z) := e^{i\vec\beta\cdot\vec\varphi(z)}.
\ea
The corresponding state is defined as  $|\vec \beta\rangle := \lim_{z\rightarrow 0}
V_{\vec\beta}(z) |0\rangle$, and satisfies
\ba
\aa k0 |\vec \beta\rangle = -i\beta_k |\vec \beta\rangle\,. 
\ea
It is a highest-weight state of the $\cW_N$ algebra:
\ba
 {\Wn r}_0 |\vec \beta\rangle &=& \Delta^{(r)}(\vec\beta)|\vec \beta\rangle,\qquad
 {\Wn r}_s| \vec \beta\rangle =0 \quad (s>0)
\label{hwc}
\ea where $\Wn r{}_s$ are the modes of the generators defined by
\begin{equation}
  {\Wn r}(z)= \sum_{s} {\Wn r}_s z^{-s-r}
\end{equation} and the eigenvalues of the zero modes are given by
\begin{eqnarray}
  \Delta^{(2)}(\vec\beta)&=&\Delta(\vec\beta)=\frac12 \vec\beta\cdot\vec\beta + iQ\vec\rho\cdot \vec\beta, \label{w2}\\
 \Delta^{(k)}(\vec\beta) &=& (-1)^k\sum_{1\leq j_1 \leq j_2\cdots\leq j_k\leq N}
\prod_{m=1}^k \left(i\vec h_{j_m}\cdot \vec\beta+Q(k-m)\right). \label{wk}
\end{eqnarray}
This condition is equivalent to the operator-product expansion
\ba
&& W^{(k)}(z) V_{\vec\beta}(0) = \frac{\Delta^{(k)}(\vec\beta)}{z^k} 
V_{\vec\beta}(0)+\cO(z^{-k+1})\,.
\ea

One important fact is that $\Delta^{(k)}(\vec\beta)$ is invariant under
the shifted action of the Weyl group given by \eqref{eq:weyl}. 
Therefore, $\Delta^{(k)}(\vec\beta)$ is given by a symmetric polynomial
of the components of $\vec\beta + iQ\vec \rho$. 

\subsection{Screening charges and null states}
To construct irreducible representations,
it is essential to understand how null states.
are produced in the Verma module over the primary state $|\vec \beta\rangle$.
A null state satisfies the highest-weight condition
(\ref{hwc}) and has vanishing inner product with all the state
in the module.  
Null states can be constructed 
by applying the so-called screening operators $\mathcal{S}^{(\pm)}_j$ to the primary 
state with a particular value of the momenta $\vec \beta'$, which we will
explain below. 

The screening operator is the integral of a special
type of the vertex operators and commutes with all the $W$ generators:
\ba
&&\mathcal{S}_j^{(\pm)}=\int \frac{dz}{2\pi i} V^{(\pm)}_j (z)=
\int \frac{dz}{2\pi i} :e^{i\alpha_\pm \vec e_j \cdot\vec\varphi(z)}:\\
&& [\Wn k_r, \mathcal{S}_j^{(\pm)}]=0\,. \label{comm}
\ea
To achieve this, we need to impose
$\Delta(\alpha_\pm \vec e_j)=1$ in particular. 
This determines  the parameters $\alpha_\pm$ to be
\begin{equation}
(\alpha_+,\alpha_-)=i(b,1/b).
\end{equation} 
Then it is easy to see that  the state
\ba
(\mathcal{S}^{(\pm)}_j)^{\ell^\pm} |\vec\beta- \ell_\pm \alpha_\pm \vec e_j\rangle\,
\ea
satisfies the highest-weight condition by using the property (\ref{comm}),
thus giving a null state in the module over $|\vec\beta\rangle$
if it is nonzero.
This state vanishes 
unless $\vec\beta$ satisfies
\ba\label{nullc}
\vec e_j \cdot \vec \beta = (1-\ell^+_j)\alpha_+ + 
(1-\ell^-_j) \alpha_-\,,\quad
(\ell^+_j, \ell^-_j=1,2,\cdots)\,,
\ea
for some $j$,
because of a nontrivial
phase in the contour integration.
If this condition is satisfied, there is a null state at level $\ell^+_j \ell^-_j$
by applying the screening charges $\mathcal{S}^{(\pm)}_j$.  

It is possible to choose $\vec\beta$ such that $\vec \beta$
satisfies the condition (\ref{nullc}) for all $j=1,\cdots, N-1$.
It is also possible that the generated null states have null states
in their own module.  
The celebrated minimal model of $\cW_N$ algebra was constructed this way.
On the other hand, the Toda theory which is relevant in our paper
has central charge $c>N-1$ and the primary fields
in general are not completely degenerate.
Therefore, we need to pay attention to
less a restrictive set of null states, which will be discussed in the next section.

\section{Level-1 null states and Young diagrams}
\label{sec:null}

\subsection{General consideration}
In the following, we will focus on the null states
which appear at level 1.  The existence of level-1 null states
implies $\ell_j^+=\ell_j^-=1$ for some $j$.  The condition (\ref{nullc})
then becomes
\begin{equation}
\vec e_j\cdot \vec \beta =0,\, \quad \text{that is,}\quad
\beta_j=\beta_{j+1}\,. \label{nullc1}
\end{equation}

More generally, let $M_{kl}$ the matrix of 
the inner products of $ W^{(k)}_0|\vec\beta\rangle$, ($k=2,\ldots,N$)
and $\vec e_l\cdot \vec a_0 |\vec\beta\rangle$ ($l=1,\ldots,N-1$).
 Then it is known that \begin{equation}
\det (M_{kl}) \propto \prod_{m<n} (\beta_n-\beta_m + i(n-m-1)Q ) \label{eq:kac}
\end{equation}
The factors on the right hand side with $n>m+1$ are the 
images of the conditions \eqref{nullc1} under the shifted Weyl action \eqref{eq:weyl}.

We consider the following conditions labeled by the Young diagram 
$Y=[l_1,l_2,\cdots,l_s]$ :
\beq
\vec\beta=(\beta_1,\cdots,\beta_N)=(\underbrace{\beta_{(1)},\cdots,\beta_{(1)}}_{l_1}, \cdots, \underbrace{\beta_{(s)},\cdots,\beta_{(s)}}_{l_s})  \label{eq:momenta}
\eeq where
\begin{equation}
\sum_{k=1}^N \beta_k=\sum_{k=1}^s k\beta_{(k)}=0. \label{eq:decouple}
\end{equation} 
The number of null states in the level-1 descendants is
$ \sum_{k=1}^s (l_k-1) $.

Let us make a few observations.  
First, the condition \eqref{eq:decouple} implies
that the null state associated with Young diagram $[N]$ 
necessarily has $\vec\beta=0$. This corresponds to the insertion of the vacuum,
and agrees with the fact that
the puncture of type $[N]$ does nothing  on the gauge theory side.
Second, the form \eqref{eq:momenta}
is exactly the same as the form of the residue of the Seiberg-Witten differential
associated to the same type of the Young diagram, \eqref{eq:residue}.
Therefore, it is natural to map the mass terms and the Toda momenta 
by identifying \eqref{eq:residue} and \eqref{eq:momenta}.
However, there is a slight problem here, because  the Weyl group
of the flavor symmetry acts linearly on the masses,
but acts on the momenta by the shifted action \eqref{eq:weyl}.
We need to take care of this discrepancy, to which we come back in \S\ref{sec:massless}.

\subsection{Explicit form of the null states}
The simplest nontrivial level-1 null state is the one 
for $N=3$ with the Young diagram $[2,1]$.
This is  the  puncture  relevant in the analysis 
in \cite{Wyllard:2009hg,Mironov:2009by}
of the correspondence between the
$\SU(3)$ gauge theory with six flavors and the Toda theory.
Therefore, let us derive the explicit form of this null state.

For $N=3$,
we have only two states $\Wn2_{-1}|\vec\beta\rangle$,
$\Wn3_{-1}|\vec\beta\rangle$ at level 1,
and a linear combination of them should become null.
To construct  level-1 states, we need only to keep
$\vec a_0$ and $\vec a_{-1}$ in the expression for
$\Wn r$.  The condition $\vec e_1\cdot \vec \beta=0$
can be solved by writing $\vec\beta=(\beta,\beta,-2\beta)$.
Then we find, after an explicit calculation,
\ba
\Wn 2_{0} |\vec\beta\rangle &=& 3\beta(\beta- i Q) |\vec\beta\rangle, \\
\Wn 2_{-1} |\vec\beta\rangle &=& i\vec\beta\cdot\vec a_{-1} |\vec\beta\rangle, \\
\Wn 3_{0} |\vec\beta\rangle &=& -2i\beta(\beta- i Q)(\beta-2i Q) |\vec\beta\rangle, \\
\Wn 3_{-1} |\vec\beta\rangle &=& (\beta-iQ)\vec\beta\cdot\vec a_{-1} |\vec\beta\rangle.
\ea
Both $\Wn2_{-1}|\vec\beta\rangle$ and $\Wn3_{-1}|\vec\beta\rangle$
are proportional to $\vec\beta\cdot\vec a_{-1}|\vec\beta\rangle$. 
Then a linear combination of them is zero. 
After the redefinition \eqref{eq:primaryW3}, the null state is given as
\ba\label{qn}
(2 \Wn2_{0} \hat W^{(3)}_{-1} - 3 \Wn2_{-1} \hat W^{(3)}_{0})|\vec\beta\rangle\approx 0\, 
\ea 
in the Verma module. 
Indeed, this state vanishes in the free-boson representation.
In this rather simple case, the null state condition can be found 
by directly studying the structure
of the Verma module as in \cite{Mironov:2009by},
but the free-boson representation is crucial to obtain the general structure.

\subsection{Physical states}
\label{sec:massless}
In the Liouville theory, the states whose momentum are of the form $\beta= p-iQ/2$ 
play a special role. 
They describe the propagation of the wave in the $\varphi$ space 
with momentum $p$, and they are the states to be inserted and integrated over
in the intermediate channel when one calculates the four-point function
by combining two three-point functions. 
On this class of states, the $L_0$ eigenvalue $p^2+Q^2/4$ is real
and bounded from below. Finally the reflection \eqref{eq:liouville-reflection}
acts on $p$ by just flipping the sign: $p\to -p$.
These features  motivated the authors of \cite{Alday:2009aq} to identify the $\SU(2)$ mass parameter $m$ with $p$ under the correspondence of the  $\SU(2)$ gauge theory and the Liouville theory.

In the Toda theory, the states whose momenta are of the form 
\begin{equation}
\vec\beta = \vec p - i Q \vec \rho \label{eq:full-physical}
\end{equation} have similar features. Namely, the $L_0$ eigenvalue is real 
and bounded from below; the shifted Weyl action on $\vec\beta$, \eqref{eq:weyl},
acts linearly on $\vec p$; and they appear in the intermediate channel when calculating four- and higher-point functions. 
These facts led  Wyllard in \cite{Wyllard:2009hg} to identify $\vec p$ with the 
$\SU(N)$ mass parameters of the gauge theory.
Let us call this class of states the physical states.

What are the physical states for a general Young diagram $Y=[l_1,\ldots,l_s]$?
The problem is that the conditions \eqref{eq:full-physical} and
\eqref{eq:momenta} are incompatible. Still, 
under the $\SU(N)$ gauge theory--Toda theory correspondence, 
one expects to integrate over a real subspace of the momenta \eqref{eq:momenta}
to get the partition function of the gauge theory. 
We propose the solution to this question below.

Let us denote by $P_Y$ the set of real vectors of the form \eqref{eq:momenta}. 
Take the formula of $L_0$, \eqref{w2}. When $\vec\beta\in P_Y + iP_Y$, 
we have \begin{equation}
L_0 = \frac12\vec\beta\cdot\vec\beta + i Q\vec \rho\cdot \vec \beta
= \frac12\vec\beta\cdot\vec\beta + i Q\vec \rho_Y\cdot \vec \beta \label{eq:L0Y}
\end{equation} where $\vec\rho_Y$ is the projection of $\vec \rho$ onto $P_Y$.
Explicitly, $\vec\rho_Y$ is given by \begin{equation}
\vec\rho-\vec\rho_Y=\vec\rho_{l_1} \oplus \vec\rho_{l_2} \oplus \cdots \oplus \vec\rho_{l_s}\label{eq:rhoY}
\end{equation} where \begin{equation}
\vec\rho_k= (\frac{k-1}2,\frac{k-3}2,\ldots,\frac{1-k}2)
\end{equation} is the Weyl vector of $A_{k-1}$, and $\oplus$ signifies that we concatenate the components of the vectors to form a vector with more components.
It is straightforward to check that $\vec\rho_Y\in P_Y$. 
Then, \eqref{eq:L0Y} becomes \begin{equation}
L_0 = \frac12(\vec \beta+i Q \vec\rho_Y) \cdot
(\vec \beta+iQ \vec\rho_Y) +\frac12 Q^2 \vec\rho_Y\cdot\vec\rho_Y.
\end{equation} 
Now it is easy to see that $L_0$ is positive definite when
the momenta is of the form \begin{equation}
\vec \beta = \vec p - i Q \vec\rho_Y, \qquad \vec p\in P_Y, \label{eq:momentaY}
\end{equation}  i.e.~when $\vec p$ has the form \begin{equation}
\vec p = (\underbrace{p_{(1)},\cdots,p_{(1)}}_{l_1}, \cdots, \underbrace{p_{(s)},\cdots,p_{(s)}}_{l_s}) .
\end{equation} 

Next, consider the shifted action of the Weyl group on $\vec\beta$.
It acts linearly on $\vec\beta + iQ \vec\rho$. 
Therefore, on the momenta \eqref{eq:momentaY}, 
it acts linearly on $\vec p + iQ (\vec \rho-\vec\rho_Y)$.
The form of $\vec\rho-\vec\rho_Y$, \eqref{eq:rhoY}, 
then means that 
the Weyl group of the flavor symmetry \eqref{eq:flavor},
which acts on $\vec\beta$ via the shifted formula \eqref{eq:weyl},
acts linearly on $\vec p$ in \eqref{eq:momentaY}.

Finally, consider the eigenvalues $\Delta^{(k)}$, \eqref{wk},
of the zero modes of the generators $\Wn k$. 
As we discussed there, they are  linear combinations 
of the symmetric polynomials of the components of $\vec\beta+ i Q \vec\rho$.
The symmetric polynomials of the components of $\vec\beta+iQ\vec\rho$ are,
in turn, polynomials with real coefficients of the power sums of the components
of $\vec\beta+iQ\vec\rho$. For the momenta of the form \eqref{eq:momentaY},
these power sums are \begin{equation}
\sum_{k=1}^N \left[\beta_k + i Q(\rho_k-\rho_{Y,k})\right]^\ell
= \sum_{k=1}^s \sum_{m=1}^{l_s} \left[p_{(k)} + i Q (\frac{l_s-1}2-m)\right]^\ell,
\end{equation} which are all real. 
Therefore,
if $\vec\beta$ is of the form \eqref{eq:momentaY},
the elementary symmetric polynomials of the components of $\vec \beta$ are real,
just as when  $\vec\beta$ is of the form \eqref{eq:full-physical}.
This property is a necessary condition to have
a unitary representation of the $\cW_N$ algebra, because the linear combinations
of the  zero modes $W^{(k)}_0$ corresponding to the elementary symmetric
polynomials should be all Hermitian.
This suggests that the state of the Toda theory
with momenta of the form \eqref{eq:momentaY} is unitary.

Combining these observations,
we propose that  the momenta of the form \eqref{eq:momentaY}
correspond to 
the mass parameters \eqref{eq:residue} of the gauge theory
associated to the same diagram $Y$
under the identification $m_k' = p_{(k)}$. 
One corollary is that the massless puncture corresponds to 
the momenta $\vec\beta =- i Q \vec\rho_Y$. 

\subsection{Null states in the semiclassical limit}
\label{sec:classical}

The correspondence of the level-1 null states and the punctures of the
Seiberg-Witten curve becomes more illuminating if we take
the limit where $Q$ is very small compared to
any component of $\vec\varphi$.  
In this limit, the action of the derivative on $\vec \varphi$ in 
(\ref{miura1}) can be neglected and we can replace the derivative
by a parameter,
\ba
Q \frac{d}{dz}\rightarrow ix\,.
\ea
We call this operation   the semiclassical limit in the following,
because it replaces the pair $(Qd/dz, z)$ which satisfies the Heisenberg
commutation relation with the pair $(x, z)$ with the Poisson bracket $\{x,z\}=1$.
$Q$ serves the role of the Planck constant $\hbar$.
Readers should bear in mind that
we still treat $\vec\varphi$ as  quantum operators
acting on the quantum Hilbert space generated from $|\vec\beta\rangle$.
We note that, in the context of integrable system, such a limit is called
``dispersionless limit" \cite{Krichever:1992sw,Takasaki:1994xh}.

The Lax operator is then replaced by its semiclassical version:
\ba
r_N|\vec \beta\rangle &=& \mathopen{:}\prod_{m=1}^N \left( x
-i\vec h_m \cdot \partial\vec \varphi(z) \right)\mathclose{:}|\vec\beta\rangle = \sum_k  \wn k(z) x^{N-k}  |\vec\beta\rangle.
\label{miura2}
\ea Here, $r_N$ and $\wn r$  are the semiclassical limits of $R_N$ and $\Wn r$
 up to powers of $i$.
We define the modes of $\wn k(z)$ as usual,
\begin{equation}
\wn k(z)= w^{(k)}_0 z^{-k} 
+ w^{(k)}_{-1} z^{-k+1} + \cdots.
\end{equation}
In the following, we use the abbreviation 
\begin{equation}
A^m(z)=i\vec h_m \cdot \partial\vec \varphi(z), \qquad
\end{equation} to simplify the formulas;
the relation  $A^m(z) |\vec \beta\rangle \sim (\beta_m/z + O(1))|\vec \beta\rangle$
can be used to evaluate $\wn r$ and $r_N$.

As we already noted, the relevant part of $A^m(z)$ in the construction
of the level-1 null state is $A^m(z) =A^m_{0} z^{-1} + A^m_{-1}+\cdots$.
To see the correspondence with the Seiberg-Witten curve,
it is convenient to consider $r'_N=  z^Nr_N$
and to introduce  $v=zx$.
The Lax operator (\ref{miura2}) becomes
\ba
r'_N|\vec\beta\rangle &=&
\label{rnz} 
 \sum_{k=0}^{N} (\wn k_0 + \wn k_{-1}z+\cdots) v^{N-k}|\vec\beta\rangle\\
&=& (w_0(v) + w_{-1} (v) z+\cdots)|\vec\beta\rangle 
\ea where 
\begin{equation}
w_s(v):= \sum_{k=1}^N  \wn k_s v^{N-k} .
\end{equation}

We note that $\wn 0(z)=1$, $\wn 1(z)=0$ because $\sum_m A^m_{s}=0$.
The explicit form of $w_0(v)$ and $w_{-1}(v)$ becomes,
after dropping the irrelevant part,
\ba
w_0(v) |\vec\beta\rangle = \prod_{m=1}^N (v-\beta_m)|\vec\beta\rangle\,,\qquad
w_{-1}(v)|\vec\beta\rangle
=w_0(v)\sum_{m=1}^N \frac{A^m_{-1} }{v-\beta_m}|\vec\beta\rangle\,.
\label{w-1}
\ea
Then, to find the null states in the semiclassical limit, 
we need to show that a linear combination of  $w^{(k)}_{-1}$ becomes null
whenever any neighboring pair of $\beta_m$ coincides.

This can be seen very easily.
Suppose $\beta_1=\beta_2=\beta$.
Then the function $w_0(v)$ has the second order zero at $v=\beta$:
\ba
w_0(v) \propto(v-\beta)^2. 
\ea
Eq.(\ref{w-1}) then implies that $w_{-1}(v)$ is zero at $v=\beta$:
\ba
w_{-1}(\beta) |\vec\beta\rangle
= \sum_{m=2}^N \beta^{N-m} w_{-1}^{(m)} |\vec\beta\rangle \approx 0.
\ea
This gives the explicit form of the null state in terms of $\wn m_{-1}$.

In a similar manner, 
we can write down the level-1 null states for 
 the singularity associated with the Young diagram 
$[l_1,l_2,\cdots, l_s]$.
Call the corresponding components of $A^m_0$
as $\beta_{(1)},\beta_{(2)},\ldots,\beta_{(s)}$.
Then we have a set of null states:
\ba
&& w_{-1}(\beta_{(k)})|\vec\beta\rangle\approx w'_{-1}(\beta_{(k)})|\vec\beta\rangle
\approx\cdots\approx
\frac{d^{l_k-2}}{d v^{l_k-2}} w_{-1}(\beta_{(k)})|\vec\beta\rangle\approx 0\label{nsc1}
\ea
for each $k$ with $l_k>1$.
It gives rise to the desired $\sum_{k=1}^s (l_k-1)$ null
states at level 1.

At this point, it is illuminating to point out a similarity
between the behavior of Lax operator near 
the singularity with the mass-deformed version of 
the Seiberg-Witten curve (\ref{eq:massiveSW}).  The null state conditions
(\ref{nsc1}
) implies that the behavior of
Lax operator $r'_N$ in the vicinity of  
$V_{\vec\beta}(0)$ is,
\ba
\langle \cdots r_N'(z) V_{\vec\beta}(0) \rangle = c_0 \prod_{k} (v-\beta_{(k)})^{l_k} 
+c_1 z \prod_k(v- \beta_{(k)})^{l_k-1}(v^{d_1}+\cdots) 
+\cdots.
\ea
The structure of the first two terms is exactly the same as
the last two terms in (\ref{eq:massiveSW}) 
if we replace  $m_k\rightarrow \beta_{(k)}$.
It implies that the behavior of $r'_N$ at $z=0$ coincides with
the Seiberg-Witten curve near the corresponding puncture.

The null states constructed above do not look like
the quantum null state for $N=3$ (\ref{qn}).  This type of
null state can be constructed when the Young diagram is
given in the form $[r,N-r]$.
Let us take $\vec\beta=(\underbrace{\beta_{(1)},\cdots,\beta_{(1)}}_r,\underbrace{ \beta_{(2)},\cdots,\beta_{(2)}}_{N-r})$
with $r\beta_1+(N-r)\beta_2=0$. 
Then $w_{0,1}(v)$ is given by 
\ba
w_0(v) |\vec\beta\rangle &=&
(v-\beta_{(1)})^r (v-\beta_{(2)})^{N-r}|\vec\beta\rangle, \nn \\
 w_{-1}(v) |\vec\beta\rangle&= &
 (\beta_{(1)}-\beta_{(2)})
(v-\beta_{(1)})^{r-1} (v-\beta_{(2)})^{N-r-1}A_{-1}|\vec\beta\rangle\,
\ea
where  
$
A_{-1} = \sum_{m=1}^r A^m_{-1}
$.
Expanding in terms of $v$, one finds
\ba
 w^{(m)}_{-1}|\vec\beta\rangle\,
=\frac{\beta_{(1)}-\beta_{(2)}}{N\beta_{(1)}\beta_{(2)}} m w_0^{(m)}A_{-1}|\vec\beta\rangle.
\ea
Therefore, we arrive at identities
which are the direct analog of the null state (\ref{qn}),
\ba\label{kanno}
(m w^{(m)}_0 w^{(n)}_{-1}-n w^{(n)}_0 w^{(m)}_{-1})|\vec\beta\rangle \approx 0\,,\quad
(m,n=2,\cdots, N)\,.
\ea
We will see in the next section that these properties also arise
naturally from the analysis of the Seiberg-Witten curve.

\section{Behavior of Seiberg-Witten curve at punctures}
\label{sec:SW}

In this section, we study the behavior of the differentials $\phi^{(k)}$ in the 
Seiberg-Witten curve close to the puncture.
We will explicitly see that  the null state
condition of $\cW$-algebra reproduces this behavior,
under the identification $\vev{\Wn k(z)}=\phi^{(k)}(z)$ in the semiclassical limit.
 
\subsection{$\SU(3)$ with 6 flavors}

In \S\ref{sec:null}, we explicitly
obtained the level-1 null state condition (\ref{qn}) 
of $\cW_3$ algebra.
To see the correspondence to the Seiberg-Witten curve, we should consider
the mass-deformed Seiberg-Witten curve (\ref{eq:massiveSW}) of the $\SU(3)$ gauge
theory with  six flavors.
This is given by
\ba
&&
(v-\tilde m_1)(v-\tilde m_2)(v-\tilde  m_3)z^2
+c_1(v^3-u^{(2)}v-u_3)z
\nt&&~~~~~~~~~~~~~~~~~~~~~~~~~~~~~~~~~~~~~
+c_0(v-m_1)(v-m_2)(v-m_3)=0\,.
\ea
As we reviewed in \S\ref{rev:SW}, 
we need to shift $v$ to eliminate $v^2$ terms to
make it of the form (\ref{eq:gc}).
Then we obtain the equation
\ba
x^3+\phi^{(2)}(z)x+\phi^{(3)}(z)=0
\ea
where $x=v/z$. 
Four punctures exist on the sphere  at $z=0,\infty,z_0,z_1$,
where $z_{0,1}$ are two solutions of $z^2 +c_1 z+ c_0=0$.
The punctures at $z=0,\infty$ are ``full'' punctures, 
while the punctures at $z=z_0,z_1$ are ``simple'' punctures.

Let us study the behavior of the curve near the simple puncture at $z=z_0$.
For that purpose, 
we take the Laurent expansion of $\phi^{(2)}(z)$ and $\phi^{(3)}(z)$ 
at $z=z_0$:
\ba
\phi^{(2)}(z)&=&
 \frac{\phi^{(2)}_{0}}{(z-z_0)^2}
+\frac{\phi^{(2)}_{-1}}{(z-z_0)^1}
+\cO((z-z_0)^0)\,,\nt
\phi^{(3)}(z)&=&
 \frac{\phi^{(3)}_{0}}{(z-z_0)^3}
+\frac{\phi^{(3)}_{-1}}{(z-z_0)^2}
+\cO((z-z_0)^{-1})\,, \label{eq:laurent}
\ea
whose explicit form can be found by specializing the general result below
in \S\ref{subsec:Nfull}.
We find the relation 
\ba \label{eq:null1}
2\phi^{(2)}_{0}\phi^{(3)}_{-1}-3\phi^{(3)}_{0}\phi^{(2)}_{-1}=0\,,
\ea
which is the same as the form  of the  null state condition of ${\cal W}_3$ algebra (\ref{qn}).
For the full punctures, no such relations exist.

\subsection{$\SU(3)\times \SU(2)$}

In the previous section, we considered the simple puncture 
of the $\SU(3)$ theory with six flavors; this simple puncture
carried the $\U(1)$ symmetry of the bifundamental hypermultiplets. 
The simple puncture also arises 
as the puncture coming from the superconformal case,
at $z=0$ or $z=\infty$.
Let us  check that the differentials $\phi^{(k)}(z)$ in this case
behave in the same manner.

Consider the quiver with the gauge group $\SU(3)\times \SU(2)$,
for which  the puncture at $z=0$ is simple.
Again, we start from the mass-deformed Seiberg-Witten curve \eqref{eq:massiveSW}.
We shift $v$ to eliminate $v^2$ terms, 
and set $x=v/z$, to make the curve into the form \eqref{eq:gc}.
To study the puncture at $z=0$, we perform the Laurent expansion at $z=0$
as in \eqref{eq:laurent}, and find
\beq
\phi_0^{(2)}&=& -\frac{(m_1-m_2)^3}{3} ,\nt
\phi_{-1}^{(2)}&=& \frac{c_1}{3c_0}(m_1^2+2m_2^2-(m_1+2m_2)M_1-3u_2^{(2)}) ,\nt 
\phi_0^{(3)}&=& \frac{2(m_1-m_2)^3}{27}, \nt
\phi_{-1}^{(3)} &=& -\frac{c_1}{9c_0}(m_1-m_2)(m_1^2+2m_2^2-(m_1+2m_2)M_1-3u_2^{(2)}).
\eeq
From this, we indeed find the same relation as Eq.\,(\ref{eq:null1}),
\beq
2\phi_0^{(2)}\phi_{-1}^{(3)}-3\phi_0^{(3)}\phi_{-1}^{(2)}=0\,.
\eeq
This shows that the Seiberg-Witten curve near all the simple punctures
satisfies the same relation, independently of its position.

\subsection{$\SU(N)$ with $2N$ flavors}
\label{subsec:Nfull}

As a generalization,
let us study the simple puncture of the $\SU(N)$ theory with $2N$ flavors.
The mass-deformed Seiberg-Witten curve in this case is given by
\beq
&&(v-\tilde m_1)(v-\tilde m_2)\cdots(v-\tilde m_N)z^2 
+c_1(v^N-u^{(2)}v^{N-2}-u^{(3)}v^{N-3}-\cdots-u^{(N)})z \nt
&&~~~~~~~~~~~~~~~~~~~~~~~~~~~~~~~~~~~~~~~~~~
+c_0(v-m_1)(v-m_2)\cdots(v-m_N)=0  \,.
\eeq
Let us denote by $z_{0,1}$ two solutions of $\Delta(z)=z^2+c_1z+c_0=0$ as before.

The full punctures are at $z=0,\infty$ and 
the simple punctures are at $z=z_0, z_1$, just as in the $\SU(3)$ case.
We take the Laurent expansion of $\phi^{(k)}(z)$ at the puncture
$z=z_0$:
\beq
\phi^{(k)}(z)=\frac{\phi_0^{(k)}}{(z-z_0)^k}+\frac{\phi_{-1}^{(k)}}{(z-z_0)^{k-1}}+\cO((z-z_0)^{k-2})\,.
\eeq
Explicit calculation leads to the result
\beq
\frac{\phi_{-1}^{(k)}}{k\phi_0^{(k)}}
&=&\frac{2a_1}{a_1z_0+b_1z_1}-\frac{2z_0-z_1}{z_0(z_0-z_1)} \nt&& 
-\frac{N}{N-1}\frac{z_0-z_1}{z_0(a_1z_0+b_1z_1)^2}\left(a_2z_0+u^{(2)}(z_0+z_1)+b_2z_1\right)\,,
\eeq
where \begin{align}
a_1&=\sum_k \tilde m_k, &
a_2&=\sum_{k<l} \tilde m_k \tilde m_l ;&
b_1&=\sum_k  m_k, &
b_2&=\sum_{k<l}  m_k  m_l .
\end{align}
Therefore, $\phi_{-1}^{(k)}/k\phi_0^{(k)}$ is independent of $k$.
This means that
\ba
k\phi^{(k)}_0\phi^{(j)}_{-1}-j\phi^{(j)}_0\phi^{(k)}_{-1}=0
\ea
is satisfied for all $j,k=2,\cdots,N$.
This is exactly the  form we found in \S\ref{sec:classical}
as the null state condition of $\cW_N$ algebra (\ref{kanno}) 
in the semiclassical limit.

\subsection{Puncture of type \protect{$[r,N-r]$} }

As a final generalization, we consider a puncture 
with the Young diagram $[r,N-r]$, to compare what we considered in 
\S\ref{sec:classical} in the Toda theory side.
The quiver which realizes this puncture is the one 
with the gauge group 
\begin{equation}
\SU(N)\times \SU(N-1)\times \cdots\times \SU(2(N-r))\times \SU(2(N-r-1))\times \cdots\times \SU(2).
\end{equation}
The mass-deformed Seiberg-Witten curve in this case is given in \eqref{eq:massiveSW},
where $s=2$, $l_1=r$ and $l_2=N-r$.
Besides the puncture at $z=0$, a full puncture is at $z=\infty$
and there are  simple punctures  at the solutions of $\Delta(z)=0$.

We perform the Laurent expansion of $\phi^{(k)}(z)$ at the puncture at 
$z=0$ of type $[r,N-r]$ as follows:
\ba
\phi^{(k)}(z)=
 \frac{\phi^{(k)}_0}{z^k}
+\frac{\phi^{(k)}_{-1}}{z^{k-1}}
+\cO(z^{-k+2})
\ea 
and we find that
\begin{equation}
\frac{\phi^{(k)}_{-1}}{k\phi^{(k)}_0} =
\frac1{r(N-r)}\frac{c_1}{c_0}\frac{r(M_1-m_1)m_1+(N-r)(M_1-m_2)m_2+N u^{(2)}_1}{(m_1-m_2)^2}
\end{equation}
is independent of $k$.
Therefore, \ba
k\phi^{(k)}_0\phi^{(j)}_{-1}-j\phi^{(j)}_0\phi^{(k)}_{-1}=0
\ea
is satisfied for all $j,k=2,\cdots,N$.
Therefore, as we mentioned in \S\ref{sec:classical},
we can surely show that Eq.\,(\ref{kanno}) is satisfied in this 
general case.

\section{Conclusions}
\label{sec:conclusions}

In this paper, we studied the structure of the null states
in the level-1 descendants of the $\cW_N$ algebra, and  found 
it to be labeled by a Young diagram with $N$ boxes.
This Young diagram controls the Toda momenta of the corresponding
primary operator insertion, and also controls the behavior of the generators $\Wn k(z)$
of the $\cW_N$ algebra close to the insertion. 

Under the correspondence between $\SU(N)$ quiver gauge theories and the
$A_{N-1}$ Toda field theory, the Young diagram labeling the primary operator
maps to the Young diagram labeling the puncture of the Riemann surface
on which $N$ M5-branes are wrapped. 
It was known that the Young diagram controls the behavior of 
the differentials $\phi^{(k)}(z)$ appearing  in the Seiberg-Witten curve
close to the puncture; we found that
the semiclassical limit of $\vev{\Wn k(z)}$ behaves exactly as $\phi^{(k)}(z)$ does,
thus giving another indication that the Toda theory gives the ``quantization'' of the 
Seiberg-Witten curve. 

We also discussed the real subspace of the allowed momenta for a given
Young diagram. In this real subspace,  the zero modes of the generators of 
the $\cW_N$ algebra, $\Wn k_{0}$, have real eigenvalues for all $k$.
Our proposal is that these special momenta describe the propagation of the waves 
along the intersection of a subset of the exponential walls of the Toda theory, and the subset is controlled by the Young diagram.

In this paper, we discussed the correspondence between the 4D gauge theory 
and the Toda field theory only at the vicinity of a puncture. 
Therefore, the obvious next step is to study the correspondence
on a whole Riemann surface with a number of punctures,
each labeled by a Young diagram. 

It should be straightforward to carry out
an analysis directly analogous 
to the ones performed in \cite{Alday:2009aq,Wyllard:2009hg,Mironov:2009by},
for general linear quiver gauge theories with $\SU$  gauge groups.
Namely, we predict that  Nekrasov's partition function
and the partition function on $S^4$ of these gauge theories 
correspond  to the conformal block of the $\cW_N$ algebra and 
the correlation function of the Toda field theory, respectively,
under the mapping we proposed in this paper.
To calculate the partition function,
we will need to integrate over  the real subspace
which  we discussed in \S\ref{sec:massless}.
The fusion rule among the primary operators labeled by 
different Young diagrams would also be  important in the matching.

It would also be interesting to understand the relation of the 
degenerate states of the Toda theory 
and the puncture of the Hitchin system, which underlies
Gaiotto's construction of the punctures \cite{Gaiotto:2009hg,Nanopoulos:2009uw}.
Another direction of the study would be to connect our analysis
with the analysis of the matrix model \cite{Dijkgraaf:2009pc,Itoyama:2009sc} 
which also gave the interpretation that the Toda field theory is the 
quantization of the Seiberg-Witten curve.

\section*{Acknowledgments}

Y. M. and Y. T. would like to thank the hospitality of 
the Yukawa Institute for Theoretical Physics where this work was initiated
 during the workshop ``Branes, Strings and Black Holes.''
Y. T.  is supported in part by the NSF grant NO. PHY-0503584, and by the Marvin L.
Goldberger membership at the Institute for Advanced Study.
Y. M. is partially supported by KAKENHI
(\#20540253) from MEXT, Japan.
S. S. is partially supported by Global COE Program ``the Physical Sciences
Frontier,'' MEXT, Japan.

\bibliographystyle{utphys}
\bibliography{bib}{}

\end{document}